\documentclass[12pt]{iopart}

\usepackage[dvips]{graphicx}

\eqnobysec

\newcommand{\beq}{\begin{equation}}
\newcommand{\eeq}{\end{equation}}
\newcommand{\ba}[1]{\begin{array}{#1}}
\newcommand{\ea}{\end{array}}
\newcommand{\bea}{\begin{eqnarray}}
\newcommand{\eea}{\end{eqnarray}}

\newcommand{\ben}{\begin{enumerate}}
\newcommand{\een}{\end{enumerate}}
\newcommand{\bit}{\begin{itemize}}
\newcommand{\eit}{\end{itemize}}
\newcommand{\bde}{\begin{description}}
\newcommand{\ede}{\end{description}}

\newcommand{\ds}{\displaystyle}

\newcommand{\sz}{\scriptsize}

\newcommand{\req}[1]{(\ref{#1})}

\begin{document}

\title{Motif-based communities in complex networks}

\author{A~Arenas$^{1,2}$\footnote{Author to whom any correspondence should be addressed},
A~Fern\'andez$^1$, S~Fortunato$^3$ and S~G\'omez$^1$}

\address{$^1$ Departament d'Enginyeria Inform\`{a}tica i Matem\`{a}tiques,
Universitat Rovira i Virgili,
Avinguda dels Pa\"{\i}sos Catalans 26,
43007 Tarragona, Spain}

\address{$^2$ Institute for Biocomputation and Physics of Complex Systems (BIFI),
Universidad de Zaragoza,
Corona de Arag\'on 42, Edificio Cervantes,
50009 Zaragoza, Spain}

\address{$^3$ Complex Networks Lagrange Laboratory (CNLL),
Institute for Scientific Interchange (ISI),
Viale S.~Severo 65,
10133 Torino, Italy}

\eads{\mailto{alexandre.arenas@urv.cat},
\mailto{alberto.fernandez@urv.cat},
\mailto{fortunato@isi.it} and
\mailto{sergio.gomez@urv.cat}}

\begin{abstract}
Community definitions usually focus on edges, inside and between the communities. However, the high density of edges within a community determines correlations between nodes going beyond nearest-neighbours, and which are indicated by the presence of motifs. We show how motifs can be used to define general classes of nodes, including communities, by extending the mathematical expression of Newman-Girvan modularity. We construct then a general framework and apply it to some synthetic and real networks.
\end{abstract}

\pacno{89.75}


\maketitle

\section{Introduction}
Modular structure in complex networks has become a challenging subject of study starting with its very definition \cite{firstnewman}. One of the most successful approaches has been the introduction of the quality function called {\em modularity} \cite{newgirvan,newanaly}, that accomplishes two goals: (i) it implicitly defines modules, and (ii) it provides with a quantitative measure to find them. It is based on the intuitive idea that random networks are not expected to exhibit modular structure (communities) beyond fluctuations.

A lot of work has been done to devise reliable techniques to maximize modularity~\cite{newfast,clauset,duch,amaral,pujol,newspect}. However, very little has been done to analyze the concept of modularity itself and its reliability as a method for community detection. To a large extent, the success of modularity as a quality function to analyze the modular structure of complex networks, relies on its intrinsic simplicity. The researcher interested in this analysis is endowed with a non-parametric function to be optimized: modularity. The result of the analysis will provide a partition of the network into communities such that the number of edges within each community is larger than the number of edges one would expect to find by random chance. As a consequence, each community is a subset of nodes more connected between them than with the rest of the nodes in the network. Recently, it has been shown that modularity is not the panacea of the community detection problem; in particular it suffers from a resolution limit that avoids grasping the modular structure of networks at low scales~\cite{fortunato}. Moreover, modularity is strongly focused on communities, so it cannot be used in general to detect groups of nodes revealed by alternative connectivity patterns. The only exception is represented by ``anti-communities'', i.e. groups of nodes with a few edges inside and many edges connecting different groups. The presence of anti-communities indicates that the network has a multipartite structure. Anti-communities could be detected by modularity minimization~\cite{newman06}, although the results are not so good, as we mention in section~\ref{secres}.


In general, detecting multipartite structure from first principles requires a definition of the classes that is quite different (in fact, opposite) with respect to standard community definitions. Let us consider bipartite networks, where nodes/actors are connected through other entities, for example collaboration in a work, attendance to an event, etc. In these specific cases, nodes of the same class (e.g.\ actors) are not directly linked, or share but a few edges, and usually some projection of the network in a subnetwork of only a class of nodes is needed for subsequent analysis. However any projection implies knowledge about the different classes of nodes. The definition of community must be generalized to deal with these cases. Doing it within a modularity-based framework requires a different formulation of modularity~\cite{rogerbip,barber}.

We remark that bipartite networks are characterized by the fact that any path with even length starting from a node of either class ends in the same class, due to the absence of internal edges in each class. So, if the two classes are $A$ and $B$ and we start from a node $i_A$ of class $A$, the first step leads to one of its neighbours, say $i_B$, which is in $B$, the next step to a neighbour $j_A$ of $i_B$, which is in $A$, and so on. In this way, paths of even length starting and ending in the same class may reveal bipartite structure, if there are many of them. On the other hand, in a graph with modular structure, there are many edges inside each module, so one expects accordingly a large number of paths between the nodes. In particular, one expects a large number of {\it cycles}, i.e. closed paths.

We deduce that short paths, or motifs, of a network, could be used to define and identify both communities and more general topological classes of nodes. Here we propose a general framework to classify nodes based on motifs. Classes will be defined based on the principle that they ``contain'' more motifs than a null model representing a randomized version of the network at study. We adopt the null model of modularity, i.e. a random network with the same degree/strength sequence of the original network, because modularity lends itself to a simple generalization, which makes calculations straightforward. We shall derive different extensions of modularity, where the building blocks will be the motifs and not just the edges, as in the original expression. After that, we shall maximize the new functions to detect the classes.


We stress that we use a modularity-based framework only as an illustrative example of how motifs could be defined to detect general node classes in networks, but in general our framework can be useful to any other method designed to detect substructure in networks. Note that the extended quality functions, that we shall introduce, also obey the principle of the resolution limit, which states that modularity will not be able to resolve substructures beyond a certain size limit, just like the original modularity~\cite{fortunato}. However this limit is now motif-dependent and then several resolution of substructures can be achieved by changing the motif.

The rest of the paper is structured as follows: in the next section we present the mathematical formalism of the generalized modularities; then, we test the framework on synthetic and real networks; finally we discuss the results obtained.

\section{Mathematical formulation of motif modularity}
The original definition of modularity by Newman and Girvan \cite{newgirvan} only deals with unweighted and undirected networks. Later on, Newman generalized it to cope with weighted networks \cite{newanaly}. In this work we start from an extension of modularity to weighted directed networks \cite{alexnjp}, which reduces to the previous one for undirected networks, and which is calculated as follows:
\beq
  Q(C) = \frac{1}{2w} \sum_{i=1}^N \sum_{j=1}^N \left(
    w_{ij} - \frac{w_i^{\mbox{\sz out}} w_j^{\mbox{\sz in}}}{2w}
    \right) \delta(C_i,C_j)\,,
  \label{qnewman}
\eeq
where $w_{ij}$ is the weight of the connection from the $i$th to the $j$th node,
$w_i^{\mbox{\sz out}}=\sum_j w_{ij}$ and $w_j^{\mbox{\sz in}}=\sum_i w_{ij}$ stand for their
output and input strengths respectively,
$2w=\sum_{ij} w_{ij}$ is the total strength of the network, $C_i$ is the index of the
community which node $i$ belongs to, and the Kronecker $\delta$ is $1$ if nodes $i$ and
$j$ are in the same community, $0$ otherwise. For undirected networks, the only change is that
$w_i^{\mbox{\sz out}}=w_i^{\mbox{\sz in}}\equiv w_i$. The larger the value of modularity,
the better the corresponding partition of the network into modules.

In the next subsections we develop the mathematical formulation of a motif modularity which generalizes the standard one in \req{qnewman}. First, the most general framework is explained, and then the formalism is applied to several classes of motifs.

\subsection{General motif modularity}
Let ${\cal M}=(V_{{\cal M}},E_{{\cal M}})$ be a {\em motif}
(connected undirected graph, or weakly connected directed graph), where
$V_{{\cal M}}$ is the set of $M$ nodes of the motif, and
$E_{{\cal M}} \subseteq V_{{\cal M}} \times V_{{\cal M}}$ is the set of its edges.

Let $\{w_{ij}\geq 0\ |\ i,j=1,\ldots,N\}$ be the weights of a
(directed or undirected) network of $N$ nodes, where $w_{ij}=0$ if there is no edge
from the $i$th to the $j$th node, and $w_{ij}\in\{0,1\}$ if the network is unweighted.
The nodes of the motif will be labeled by the indices
$i_1$, $i_2$, \ldots, $i_M$, all of them running between 1 and $N$.

Given a certain partition $C$ of an unweighted network in communities, the number of motifs fully included within the communities is given by
\beq
  \Psi_{{\cal M}}(C) = \sum_{i_1=1}^N \sum_{i_2=1}^N \cdots \sum_{i_M=1}^N
  \prod_{(a,b)\in E_{{\cal M}}} w_{i_a i_b}\,\delta(C_{i_a},C_{i_b})\,.
  \label{psic}
\eeq
Degenerated motifs, i.e.\ those where some nodes are counted more than once, are included in this sum. The formula also holds for weighted networks, which can be inferred from the mapping between weighted networks and unweighted multigraphs~\cite{newanaly}.

The maximum value of $\Psi_{{\cal M}}(C)$ corresponds to the partition in a single community containing all the nodes:
\beq
  \Psi_{{\cal M}} = \sum_{i_1=1}^N \sum_{i_2=1}^N \cdots \sum_{i_M=1}^N
  \prod_{(a,b)\in E_{{\cal M}}} w_{i_a i_b}\,.
\eeq

For a random network preserving the nodes' strengths, these quantities are respectively
\beq
  \Omega_{{\cal M}}(C) = \sum_{i_1=1}^N \sum_{i_2=1}^N \cdots \sum_{i_M=1}^N
  \prod_{(a,b)\in E_{{\cal M}}}
  w_{i_a}^{\mbox{\sz out}} w_{i_b}^{\mbox{\sz in}}\,
  \delta(C_{i_a},C_{i_b})
  \label{omegac}
\eeq
and
\beq
  \Omega_{{\cal M}} = \sum_{i_1=1}^N \sum_{i_2=1}^N \cdots \sum_{i_M=1}^N
  \prod_{(a,b)\in E_{{\cal M}}}
  w_{i_a}^{\mbox{\sz out}} w_{i_b}^{\mbox{\sz in}}\,.
\eeq

Now, by analogy with the standard modularity, we define the {\em motif modularity} as the fraction of motifs inside the communities minus the fraction in a random network which preserves the nodes' strengths:
\beq
  Q_{{\cal M}}(C) =
    \frac{\Psi_{{\cal M}}(C)}{\Psi_{{\cal M}}} -
    \frac{\Omega_{{\cal M}}(C)}{\Omega_{{\cal M}}}\,.
\eeq

The introduction of {\em nullcase weights} $n_{ij}$, {\em masked weights} $w_{ij}(C)$ and {\em masked nullcase weights} $n_{ij}(C)$,
\begin{eqnarray}
  n_{ij}    & = & w_i^{\mbox{\sz out}} w_j^{\mbox{\sz in}}\,, \\
  w_{ij}(C) & = & w_{ij} \delta(C_i,C_j)\,, \\
  n_{ij}(C) & = & n_{ij} \delta(C_i,C_j)\,,
\end{eqnarray}
allows the simplification of the previous expressions, in particular motif modularity:
\beq
  Q_{{\cal M}}(C) =
    \frac{\ds
      \sum_{i_1 i_2 \cdots i_M}\ \prod_{(a,b)\in E_{{\cal M}}} w_{i_a i_b}(C)
    }{\ds
      \sum_{i_1 i_2 \cdots i_M}\ \prod_{(a,b)\in E_{{\cal M}}} w_{i_a i_b}
    }
    -
    \frac{\ds
      \sum_{i_1 i_2 \cdots i_M}\ \prod_{(a,b)\in E_{{\cal M}}} n_{i_a i_b}(C)
    }{\ds
      \sum_{i_1 i_2 \cdots i_M}\ \prod_{(a,b)\in E_{{\cal M}}} n_{i_a i_b}
    }\,.
  \label{qmotif}
\eeq

Motif modularity may be further generalized by relaxing the condition that all nodes of the motif should be fully inside the modules. This is done just by removing some of the maskings in \req{qmotif} as required, and possibly with the addition of some Kronecker $\delta$ functions between non-adjacent nodes of the motif. In this way, it is possible to define classes of nodes different from communities, as we shall see in subsection \ref{paths}.

\subsection{Cycle modularity}
Among the simplest possible motifs, triangles are the ones which have deserved more attention in the networks literature. For instance, it has been shown that real networks have higher clustering coefficients than expected in random networks \cite{clust}. Thus, it would be desirable to be able to find ``communities of triangles''. Our approach consists in the definition of a {\em triangle modularity} $Q_{\triangle}(C)$, based on the triangular motif
$E_{\triangle}=\{(1,2),(2,3),(3,1)\}$, which reads:
\beq
  Q_{\triangle}(C) =
    \frac{\ds
      \sum_{ijk} w_{ij}(C) w_{jk}(C) w_{ki}(C)
    }{\ds
      \sum_{ijk} w_{ij} w_{jk} w_{ki}
    }
    -
    \frac{\ds
      \sum_{ijk} n_{ij}(C) n_{jk}(C) n_{ki}(C)
    }{\ds
      \sum_{ijk} n_{ij} n_{jk} n_{ki}
    }\,.
\eeq

Triangle modularity is trivially generalizable to cycles of length $\ell$, making use of the cyclical motif
$E_{{\cal C}^{(\ell)}}=\{(1,2),(2,3),\ldots,(\ell-1,\ell),(\ell,1)\}$. The number of these motifs within the communities is given by
\beq
  \Psi_{{\cal C}^{(\ell)}}(C) =
    \sum_{i_1 i_2 \cdots i_{\ell}}
    w_{i_1 i_2}(C) w_{i_2 i_3}(C) \cdots w_{i_{\ell-1}i_{\ell}}(C) w_{i_{\ell}i_1}(C)\,.
\eeq
The full formula for the cycle modularity $Q_{{\cal C}^{(\ell)}}(C)$ follows immediately from it.

If the network is directed, other non-cyclical motifs exist. We skip them, since their derivation is straightforward.

\subsection{Path modularity \label{paths}}
A {\em path} ${\cal P}^{(\ell)}$ of length~$\ell$ is simply the linear motif
$E_{{\cal P}^{(\ell)}}=\{(1,2),(2,3),\ldots,(\ell,\ell+1)\}$.
We remark that cycles are closed paths, but here we shall only consider open paths. The number of paths of length~$\ell$ fully inside the communities is given by
\beq
  \Psi_{{\cal P}^{(\ell)}}(C) =
    \sum_{i_1 i_2 \cdots i_{\ell+1}}
    w_{i_1 i_2}(C) w_{i_2 i_3}(C) \cdots w_{i_{\ell} i_{\ell+1}}(C)\,.
\eeq
Note that this expression equals the sum of the components of the $\ell$th power of the masked weight matrix.

The path of length~$\ell=1$ corresponds to the simplest motif $E_{{\cal P}^{(1)}}=\{(1,2)\}$, which is just a single edge, so its motif modularity \req{qmotif} equals the standard definition of modularity \req{qnewman}.

Paths of length~$2$ are also useful for the analysis of bipartite networks, provided one removes the constraint that all nodes of the path belong to the same module. If one allows that the middle node of a path of length $2$ could be any node of the network, whereas the first and third nodes are kept within the same group, the path can be used to discover relationships between nodes of different groups. If a network is bipartite, for instance, there will be many paths of length $2$ starting from a class and returning to it from the other class. If only the extremes of the path $\tilde{{\cal P}}^{(\ell)}$ are required to be inside the community, their total number is given by
\beq
  \Psi_{\tilde{{\cal P}}^{(\ell)}}(C) =
    \sum_{i_1 i_2 \cdots i_{\ell+1}}
    w_{i_1 i_2} w_{i_2 i_3} \cdots w_{i_{\ell} i_{\ell+1}} \delta(C_{i_1},C_{i_{\ell+1}})\,.
\eeq
In this case, the calculation makes use of the $\ell$th power of the weight matrix (instead of the masked weight matrix), and the masking is applied to the sum of their components.

\section{Examples and tests \label{secres}}
When one is faced with the problem of community detection in a particular network, the first thing to do should be to answer the following question: what sort of connectivity patterns or motifs are pertinent in this study? According to the answer, it is straightforward to select one of the possible motif modularities. We present in this section examples of the application of the previous framework to two synthetic networks. Finally, we perform two tests on real networks for which the real partitions observed are known.

\begin{figure}[!tpb]
  \begin{center}
  \begin{tabular}[t]{cc}
    \multicolumn{1}{l}{(a)}
    &
    \multicolumn{1}{l}{(b)}
    \\ \\
    \mbox{\includegraphics*[width=.57\textwidth]{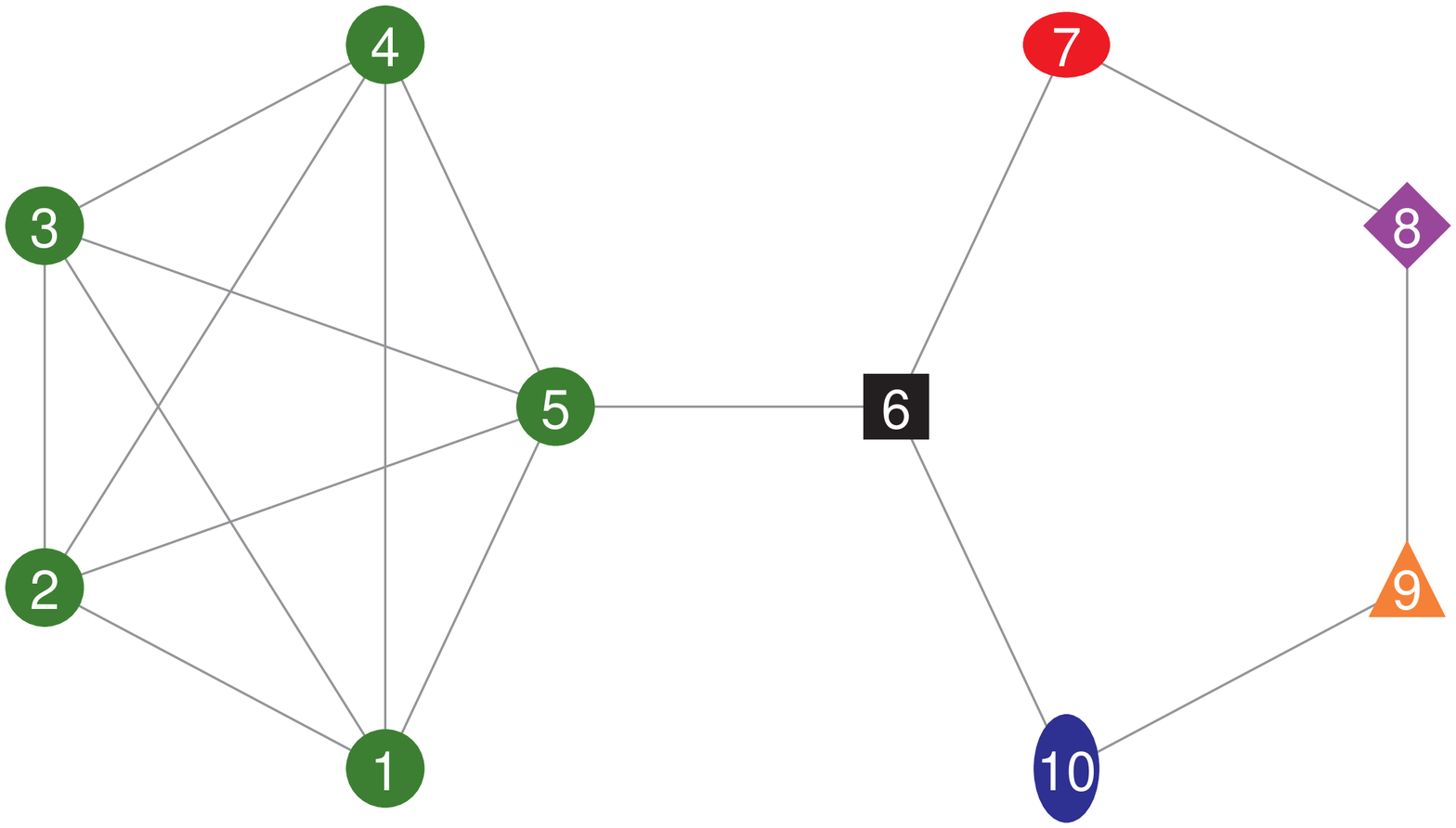}}
    &
    \mbox{\includegraphics*[width=.33\textwidth]{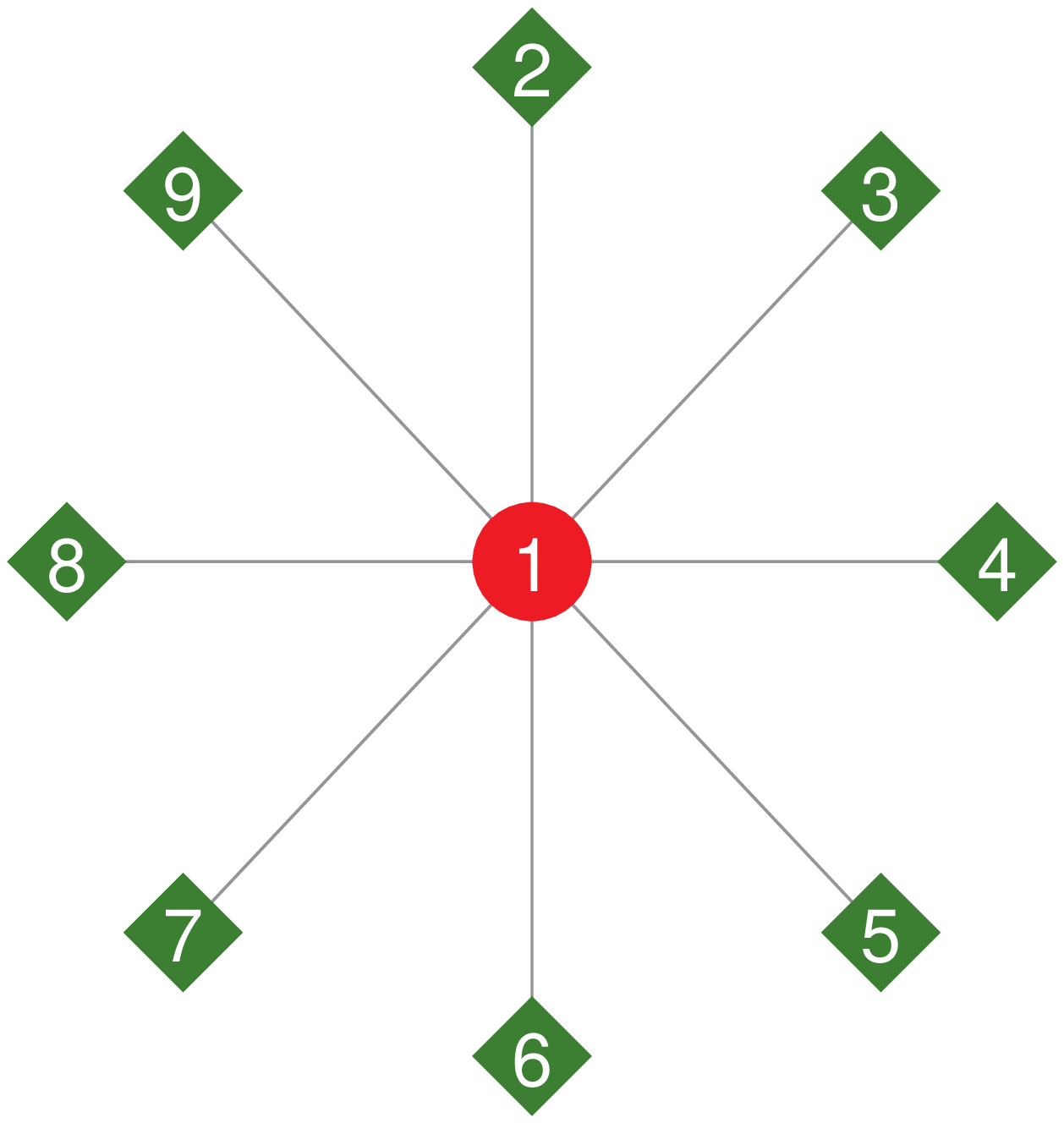}}
  \end{tabular}
  \end{center}
  \caption{Results for two synthetic networks: (a) Clique~\& circle network, with triangle modularity; (b) Star network, with paths of size 2 modularity with free intermediate node (see text for details). Members of the same class are depicted using equal symbol and color.}
  \label{netstoy}
\end{figure}

The synthetic networks that we have generated for this purpose are the clique~\& circle network and the star network. In figure~\ref{netstoy} we show these networks as well as the classes found using different motif modularities. Suppose we want to find node classes by means of triangles. When we optimize the triangle modularity for the clique~\& circle network, the clique forms a community whereas the nodes of the circle are separated into five singleton communities. This is due to the absence of triangles within the circle. On the contrary, the standard modularity identifies the circle as a community.

The second example, the star network, is a case where the path motifs prove to be useful. This network can be seen as a simple bipartite network with eight actors (the leaf nodes) and just one event (the hub node). In this case, recalling what we have said in the previous section, the path modularity of length~$2$ with a free intermediate node is the proper motif modularity to use. The results confirm that the star is decomposed in two classes, one for the leaves and another for the hub. The same partition is obtained for any even path length with free intermediate nodes, while for odd path lengths all nodes are joined in a single community. This holds as well if one maximizes the standard modularity; however, the correct partition of the network can be recovered by modularity minimization.

\begin{figure}[!tpb]
  \begin{center}
  \begin{tabular}[t]{cc}
    \multicolumn{1}{l}{(a)}
    &
    \multicolumn{1}{l}{(b)}
    \\ \\
    \mbox{\includegraphics*[width=.50\textwidth]{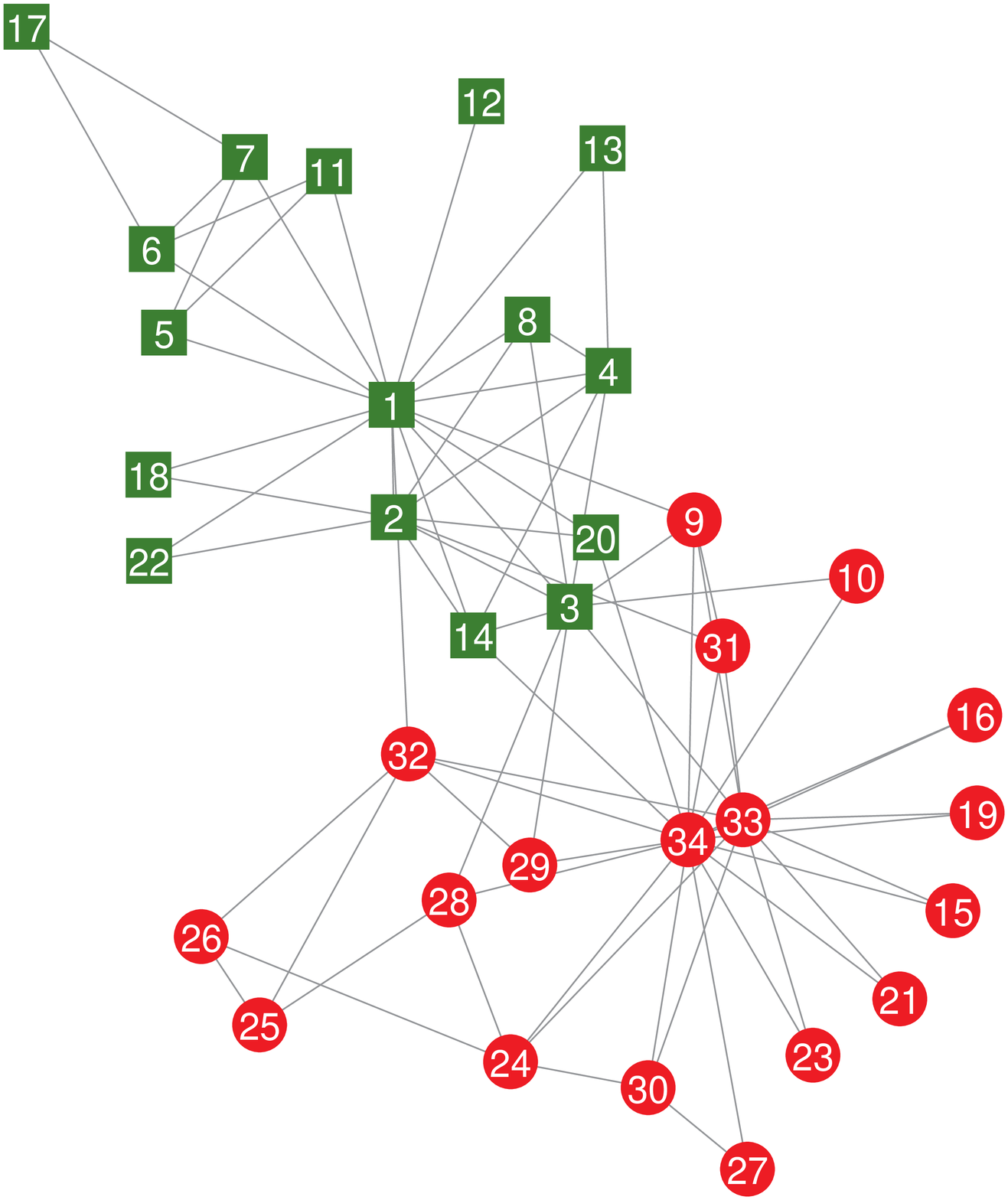}}
    &
    \begin{tabular}[b]{c}
    \mbox{\includegraphics*[width=.40\textwidth]{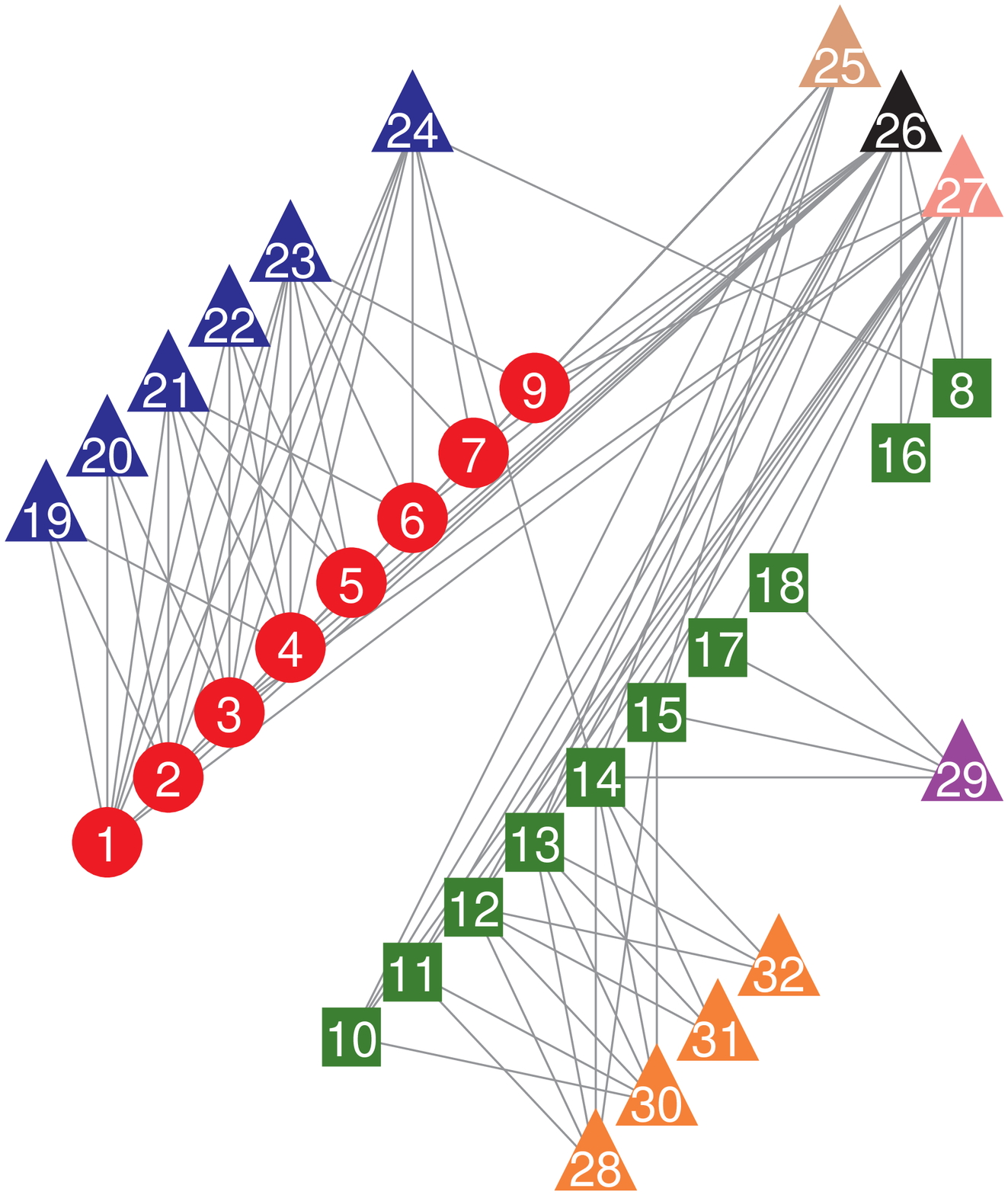}}
    \\ \mbox{\rule{0pt}{60pt}}
    \end{tabular}
  \end{tabular}
  \end{center}
  \caption{Results for two real networks: (a) Zachary Karate Club network. We depict the real splitting obtained when using several path and cycle modularities; (b) Southern Women Event Participation network. We depict the results of the analysis of this multipartite network without any projection, simply applying modularity of path free intermediate of length $2$. Remarkably the results show clearly the role differentiation of women and events, as well as the splitting of women according to the events participation that has been reported in the literature.}
  \label{netsreal}
\end{figure}

The real networks used for testing are the Zachary Karate Club network \cite{zachary} and the Southern Women Event Participation network \cite{davis,freeman}. A description of each network can be found in their respective references. For the mathematical analysis presented here the interesting fact regarding these networks is that we know the real splittings occurred in the Zachary network, as well as the most plausible classification assigned in the literature to the Women Event Participation data, as reported by Freeman \cite{freeman}. In figure~\ref{netsreal} we show both networks as well as their respective partitions.

For the Zachary Karate Club network, the nature of the data suggests to try an optimization of path modularities, since the decision of following any of the two leaders during the splitting of the club surely depended on higher order friendship relationships (friends of friends, and so on). When a path modularity of length~1 is considered (i.e.\ the classical definition of modularity), the best partition obtained splits each one of the two real communities into two sub-communities, yielding a partition in four communities. But when one looks for a more compact structure of the communities, which can be accomplished by increasing the length of the paths, the optimization of path modularity delivers the real splitting observed, for all path lengths we have used (from~$2$ to~$6$). The same result is obtained when the paths are replaced by cycles (lengths from~$4$ to~$9$). Triangles give almost the exact partition, but with two exceptions: nodes~$10$ and~$12$ become isolated, because they do not belong to any triangle.

The second network tested is a multipartite network. In this case, as well as for the star network, the use of path modularity of length~$2$ with a free intermediate node is crucial, and it accounts for the role differentiation between women and events. The results not only reveal the two roles of events and women, but also recover their internal split according to their participation in events, a classification made by social scientists \cite{freeman} (with the same exception of one woman, as in the weighted projection and bipartite methods in \cite{rogerbip}). In this case, the minimization of standard modularity is only able to separate women and events, with no further subdivision.

\section{Conclusions}
In this work we have shown that a general classification of node groups in networks is possible if one uses motifs as elementary units, instead of simple edges. To show that, we generalized Newman-Girvan modularity by replacing edges with motifs. The new versions of modularity obtained have been tested on synthetic and real networks, and are able to recover expected connectivity patterns in networks, both when the networks have modular structure and when they have multipartite structure. However, the principle goes beyond the use of modularity and could inspire promising alternative frameworks.

\subsection*{Acknowledgments}
This work has been supported by Spanish Ministry of Science and Technology Grant FIS2006-13321-C02-02.


\section*{References}
\begin{thebibliography}{99}
\bibitem{firstnewman}
Girvan M and Newman M E J 2002 Community structure in social and biological networks {\em Proc. Natl. Acad. Sci. USA} {\bf 99} 7821

\bibitem{newgirvan}
Newman M E J and Girvan 2004 Finding and evaluating community structure in networks {\em Phys. Rev. E} {\bf 69} 026113

\bibitem{newanaly}
Newman M E J 2004 Analysis of weighted networks {\em Phys. Rev. E} {\bf 70} 056131

\bibitem{newfast}
Newman M E J 2004 Fast algorithm for detecting community structure in networks {\em Phys. Rev. E} {\bf 69} 066133

\bibitem{clauset}
Clauset A, Newman M E J and Moore C 2004 Finding community structure in very large networks {\em Phys. Rev. E} {\bf 70} 066111

\bibitem{duch}
Duch J and Arenas A 2005 Community identification using Extremal Optimization {\em Phys. Rev. E} {\bf 72} 027104

\bibitem{amaral}
Guimer\`a R and Amaral L A N 2005 Functional cartography of metabolic networks {\em Nature} {\bf 433} 895

\bibitem{pujol}
Pujol J M, B\'ejar J and Delgado J 2006 Clustering Algorithm for Determining Community Structure in Large Networks {\em Phys. Rev. E} {\bf 74} 016107

\bibitem{newspect}
Newman M E J 2006 Modularity and community structure in networks {\em Proc. Natl. Acad. Sci. USA} {\bf 103} 8577

\bibitem{fortunato}
Fortunato S and Barth{\'e}lemy M 2007 Resolution limit in community detection {\em Proc. Natl. Acad. Sci. USA} {\bf 104} 36

\bibitem{newman06} Newman M E J 2006 Finding community structure in networks using the eigenvectors of matrices {\em Phys. Rev. E} {\bf 74} 036104

\bibitem{rogerbip}
Guimer\`a R, Sales-Pardo M, Amaral L A N 2007 Module identification in bipartite and directed networks {\em Phys. Rev. E} {\bf 76} 036102

\bibitem{barber}
Barber M J 2007 Modularity and community detection in bipartite networks {Preprint} arXiv:0707.1616

\bibitem{alexnjp}
Arenas A, Duch J, Fern\'andez A and G\'omez S 2007 Size reduction of complex networks preserving modularity {\em New J. Phys.} {\bf 9} 176

\bibitem{clust}
Milo R, Shen-Orr S, Itzkovitz S, Kashtan N, Chklovskii D and Alon U 2002 Network Motifs: Simple Building Blocks of Complex Networks {\em Science} {\bf 298} 824

\bibitem{zachary}
Zachary W W 1977 An information flow model for conflict and fission in small groups {\em J. Anthr. Res.} {\bf 33} 452

\bibitem{davis}
Davis A, Gardner B B and Gardner M R 1941 {\em Deep South} (Chicago: The University of Chicago Press)

\bibitem{freeman}
Freeman L 2003 Finding Social Groups: A Meta-Analysis of the Southern Women Data {\em Dynamic Social Network Modeling Analysis: Workshop Summary and Papers} ed R Breiger, K Carley and P Pattison (Washington DC: The National Academies Press) p 39
\end{thebibliography}
\end{document}